\documentclass[11pt]{article}

\usepackage{graphicx}
\usepackage{xcolor}
\usepackage{multirow}
 \usepackage{amsmath} 
\usepackage{mathtools}
\usepackage[text={6.5in,9.5in},centering]{geometry}

\begin{document}

\begin{center}
{\large{From the Division of Geophysics, Meteorology and Geomagnetism:}}\\ \vskip 2 truemm
{\large{Viktor F. Hess\index{Hess, Victor} (Wien),\\MEASUREMENTS OF THE PENETRATING RADIATION\\DURING SEVEN BALLOON FLIGHTS\\
{\em (Translated and commented by  A. De Angelis and C. Arcaro b. Schultz)}
}}

\end{center}
\vskip 1cm

In the previous year I already had the opportunity to undertake two balloon flights for the 
exploration of penetrating radiation: about the first flight I have already
reported at the  {Naturforscherversammlung  meeting} in Karlsruhe.\footnote{{This journal}  12, 998-1001; Wien  {Session Reports} 120, 1575-1585, 1911.} Both trips revealed no fundamental change in radiation compared to the values observed on the ground up to an altitude of 1100 meters. Consistently, Gockel\footnote{This journal 12, 595-597, 1911.} failed in two balloon ascents to find the expected reduction of the radiation with height. It was hence concluded that there had to be another source of penetrating radiation besides
$\gamma$-radiation from radioactive substances in the Earth's crust.\\
\indent A subsidy from the Royal Academy of Sciences of Vienna has allowed me now to perform a series of another seven balloon ascents this year, from which  larger and in many ways more extended observational material was obtained.\\
\indent In the first place, two Wulf radiation detectors with three-millimeter thick walls, perfectly sealed and able to withstand all pressure variations during the ascents, were used for the observation of the penetrating radiation.\\
\indent Instrument 1 had an ionization volume of 2039 cubic centimeters, and its capacity was 1.597 cm. Instrument 2 had a volume of 2970 cm$^3$, the capacity was equal
to 1.097 cm.  Thus, a charge loss of 1 volt per hour corresponded to an ionization rate of $q  = 1.56$ ions/cm$^3$/s in instrument 1, and to $q  = 0.7355$ ions per cm$^3$ and per second in instrument 2.\\
\indent Both instruments were electrolytically galvanized inside
to reduce as much as possible the radiation from the container walls. This suggestion has been made by   Priv.-Doz. Dr. Bergwitz. After this treatment, device 1 indicated a normal ionization of about 16 ions, and device 2 one of about 11 ions per cm$^3$ per second. The company G\"unther \& Tegetmeyer in Braunschweig applied another fundamental improvement to the instruments: until then the focusing on the fibers {\em [Translator's note: these are the metalized silicon glass wires replacing the traditional metal leaves in Wulf's electroscope]} was carried out by exclusively moving the eyepiece, which involved a considerable change in the magnification and caused in repeated adjustments differences up to 0.5 in readings. Now they mounted a sliding concave lens in the eyepiece tube, which manages the focusing with different positions of the fibers without noticeable magnification variations. Hereby, the setting accuracy is greatly improved.\\
 \indent Since the thickness of the walls of the instruments 1 and 2 was of three millimeters, essentially only $gamma$ rays could be effective. {\em [Translator's note: in 1912 only $\alpha$, $\beta$ and $\gamma$ radioactivity were known, while the penetrating power of protons was yet unknown]}
\indent In order to study simultaneously the behavior of the $\beta$ rays, I further used a third instrument: this device was not made air-tight, but consisted in an ordinary Wulf   two-fiber electrometer over which a cylindrical ionization vessel  of a volume of 16.7 liters was put, made of the slimmest zinc sheet commercially available (wall thickness of 0.188 mm), so that soft rays with the characteristics of $\beta$ rays could also play an effective role. A 20 cm high zinc spike fixed to the fiber support acted as a charge disperser. The capacity was 6.57 cm.

The loss of insulation in the thick-walled Wulf radiation detectors 1 and 2 was determined as usual with a lowered guard tube. The hourly charge loss was 0.2 V for instrument 1, and 0.7 V for instrument 2. An insulation fault due to very damp weather has never occurred.\\
\indent  For the determination of the loss of insulation of the thin-walled instrument the ionization chamber needs to be removed. The voltage loss of the electrometer alone is then subtracted from the observed total dispersion, after conversion to the capacity of the fully assembled instrument.

 Since all observers of the penetrating radiation on towers had  noted a decrease in the radiation, while Gockel and I  had not been able so far to find such decrease with certainty in  free balloon flights, it was especially important to carry out measurements during longer flights at lower altitude obtain to safe average values. Parallel observations with the thin-walled instrument 3 were supposed to show whether the softer radiation behaves as the actual $\gamma$ radiation.\\
\indent Special attention was paid to the radiation fluctuation. Pacini\footnote{Le Radium 8, 307-312, 1911.} has found during parallel observations of two Wulf radiation detectors at one-hour reading intervals unequivocal simultaneous fluctuations in the rate of discharge both on land and over the sea; the cause of the fluctuations obviously lies outside the instruments, in the radiation itself. It was now very important to ascertain whether such simultaneous changes of the radiation in several instruments are also noticeable in the balloon. Since this measurement is most impeccable during continuous flights at the same height, I have carried out the majority of observations at night flights.

The last and most important point of this investigation was the measurement of the radiation in the highest possible heights. While in the six trips made from Vienna the low gas carrying capacity of the balloon and meteorological conditions did not allow this, I managed to carry out measurements up to 5350 m in an ascent made with hydrogen from Aussig at the Elbe.\\
\indent Before each trip, several hours of control observation were made with all three devices. Here, the devices were attached  to the balloon basket in the same way during the flight itself. The observations before the ascents were mad at the club field of the imperial-royal Austrian Aeroclub, a leveled grass pitch  in the Prater in Vienna. King\footnote{Phil. Mag. (6) 23, 242, 1912.} has recently conjectured that  balloon observations may be disturbed by the proximity of possibly low-level radioactive ballast sand. I have never found an increase in radiation in the immediate vicinity of large ballast sand supplies.

 For the instruments 1 and 2, the air density within the ionization chamber is always the same as at the ascent place (750 mm in average). The situation is different of the thin-walled instrument 3, for which the pressure is the same as for the environment. It is therefore necessary, especially for observations at higher altitudes, to reduce the directly observed measurements. Given the proportionality of the ionization produced by the penetrating radiation with the pressure, the  observed radiation value was multiplied by the ratio of the normal pressure of 750 mm to the mean pressure $b$  during the observation interval. It must be noted that this reduction holds some uncertainty: for it is implicitly assumed that the residual radiation of the vessel walls changes proportionally with the air pressure in the vessel, which definitely does not need to be the case if this residual radiation has only weak penetration force such as for the $\alpha$ rays. Therefore, in particular for the measurements at higher altitudes the unreduced values obtained with the instrument 3 will also be discussed.\\
\indent In the tables below, $q_1$, $q_2$, $q_3$ represent the penetrating radiation in ions per cm$^3$ per second observed in  instruments 1, 2 and 3. The elementary charge is  assumed to be $e = 4.65 \cdot 10^{-10}$ esu.\\
\indent The mean height of the balloon during the relevant observation intervals (usually 1 hour each) was taken from the barograph curve by a graphic procedure. From the sea level of the respective places overflown an average value for the relative height was calculated. The hours of the day are indicated in the tables with 24-hour division.\\
\indent The detailed report on all observations made in the balloon has been handed over to the Imperial Academy of Sciences in Vienna and appears in their session reports. Here, I only want to report in detail on the two main flights and be content with the quotation of the average values for the others.
 
\newpage

\begin{center}{1$^{\mathrm{st}}$ Flight.}\end{center}
\indent This flight took place on April 17, 1912, on the occasion of a solar eclipse, which was very partial in lower Austria. From 11 am to 1 pm  observation were made at 1900-2750 m of absolute altitude over a nearly completely cloudy sky. There was no noticeable reduction in penetrating radiation as eclipsing increased. Instrument 2 for example showed before the ascent  an ionization of 10.7 ions, in 1700 m of average relative height 1 ion, thereafter in 1700 to 2100 m during the first stages of the eclipse 14.4, later at about 50 percent solar coverage 15.1 ions. Further measurements were not possible because the balloon was forced to descend to due to the cooling down of the gas.\\
\indent Thus, an increase in the radiation at about 2000 m was found. Since no influence of the eclipse on the penetrating radiation was noted, we  conclude that even if a part of the radiation should be of cosmic origin, it hardly emanates from the Sun, at least as long as one has in mind a direct, rectilinear propagating $\gamma$-radiation. This view is further corroborated by the fact that during the later journeys in  balloon I never found a pronounced difference of radiation between  day and  night.\\

\begin{center}{2$^{\mathrm{nd}}$ Flight (April 26 - 27, 1912).}\end{center}
\begin{center}
\tiny
\begin{tabular}{c|c|c|c|c|c|c|cll}
\multicolumn{4}{l}{Balloon: ``Excelsior'' (1600 cbm  illuminating gas)}&\multicolumn{6}{r}{Driver: Captain W. Hoffory.}\\
\multicolumn{4}{l}{}&\multicolumn{6}{r}{Observer: V. F. Hess.}\\\hline\hline
 &&\multicolumn{2}{c}{Average altitude}&\multicolumn{6}{|c}{Observed radiation}\\
 \cline{5-10}
 No.&Time&\multicolumn{2}{c|}{}&Instrument 1&Instrument 2&\multicolumn{3}{c}{Instrument 3}\\
  \cline{3-4}
 &&absolute&relative&&\\ 
  \cline{5-10}
   &&m&m&$q_1$&$q_2$&$q_3$&$q_3$ (reduced)\\\hline\hline
   1&16$^\mathrm{h}$ 40 - 17$^\mathrm{h}$ 40 &156&0&15.6&11.5&-&-&\multirow{5}{*}{{\huge\} }}&measured\\
     2&17$^\mathrm{h}$ 40 - 18$^\mathrm{h}$ 40 &156&0&18.7&11.8&21.0&21,0&&before the\\
      3&18$^\mathrm{h}$ 40 - 21$^\mathrm{h}$ --- &156&0&17.8&11.6&19.5&19,5&&ascent\\
 4&21$^\mathrm{h}$ 30 - 22$^\mathrm{h}$ 30 &156&0&17.8&11.3&20.0&20.0&&at the club field\\\
 5&23$^\mathrm{h}$ 26 - 0$^\mathrm{h}$ 26 &300&140&14.4&9.6&19.4&19.8&&(Vienna)\\
 6&0$^\mathrm{h}$ 26 - 1$^\mathrm{h}$ 26 &350&190&16.2&9.9&17.4&17.9&\\
 7&1$^\mathrm{h}$ 26 - 2$^\mathrm{h}$ 26 &300&140&14.4&10.1&17.7&18.1&\\
 8&2$^\mathrm{h}$ 26 - 3$^\mathrm{h}$ 32 &330&160&15.0&9.6&18.2&18.7&\\
 9&3$^\mathrm{h}$ 32 - 4$^\mathrm{h}$ 32 &320&150&14.4&9.8&18.5&19.0&\\
 10&4$^\mathrm{h}$ 32 - 5$^\mathrm{h}$ 35 &300&70&17.2&13.2&20.6&21.0&\\
 11&5$^\mathrm{h}$ 35 - 6$^\mathrm{h}$ 35 &540&240&17.8&11.8&19.6&20.8&\\
 12&6$^\mathrm{h}$ 35 - 7$^\mathrm{h}$ 35 &1050&800&17.6&10.0&18.1&20.3&\\
 13&7$^\mathrm{h}$ 35 - 8$^\mathrm{h}$ 35 &1400&1200&12.2&8.8&17.3&20.3&\\
 14&8$^\mathrm{h}$ 35 - 9$^\mathrm{h}$ 35 &1800&1600&17.5&10.9&17.3&21.3&\\
\end{tabular}

\end{center}

\indent The table above gives an overview of all observations during the second flight. The ascent took place at 11 pm. It was possible
by careful guidance to hold the balloon at almost the same altitude (300-350 m) for 6 hours, which was important for determining the variations in radiation. The flight led from the Prater to the South then calm arrived and finally the balloon drifted towards North over Florisdorf, Stockerau, Guntersdorf to M\"ahren. At 10$^\mathrm{h}$30 am, after reaching a maximum altitude of 2100 m we landed in Pausram south of Br\"unn. The sky was cloudless throughout the entire 11 hour flight.\\
\indent Above all, the table shows that at low altitude above the ground, the radiation is actually lower than on the ground itself. Let us build the averages:
\begin{center}
\tiny
\begin{tabular}{llll}
&Instrument 1&Instrument 2& Instrument 3\\
Before&&&\\
ascent&$q_1=17.5$&$q_2=11.55$&$q_3=q_\mathrm{red.}=20.2\frac{\mathrm{Ions}}{\mathrm{cm^{3}\,s}}$\\
\\
At 140 to&&&\\
190 m above&&&\\
ground&$q_1=14.9$&$q_2=9,8$&$q_3=18.2\,q_\mathrm{3red.}=18.7$
\end{tabular}
\end{center}
\indent Accordingly, the ionization differences are 2.6, 1.8, 2.0 and 1.5. On average, the difference is about 2 ions. This decrease of the radiation by 2 ions apparently originates from the absorption of the $\gamma$-radiation of the radioactive substances of the Earth's crust in the air. According to King's calculation\footnote{Phil. Mag. (6) 23, 247, 1912.}, at 160 m the $\gamma$-radiation is already weakened to 24\%. The above-mentioned difference of 2 ions thus corresponds to about 3/4 of the total ionization strength, which is generated by the radioactive $\gamma$-radiation of the Earth's crust. The total $\gamma$-radiation of the Earth's crust is therefore likely to produce about 3 ions per 1 cm$^3$ and second in zinc vessels.  {At the low height of 160 m, the decrease in the induction  and the possible increase in radiation from above may be ignored.}

 The fluctuations of the radiation are also very evident in the table. First of all, we can already see from the observations on the ground that the information given by several instruments placed next to each other is not exactly simultaneous. This is primarily  caused by  reading errors.
 
 The hourly decrease of the fiber distance was on average of 6 scale divisions for instrument 1,  9 scale divisions for  instrument 2, 15 scale divisions
for instrument 3. If we take 0.1 scale units as possible error, 0.2 scale units in the extreme case, then an error of 0.4, and in extreme cases 0.8 division marks, may arise at the initial and the final reading of both fibers. This gives as a possible error on reading at hourly intervals 7\% for instrument 1, 
4.4\% for instrument 2, 2.7\% for instrument 3  (in extreme cases, 14\%,  9\% and 5\%, respectively).

We will be able to address only those changes with certainty as a real fluctuation of radiation, which are given at the same time in all instruments and which exceed in size the possible errors in the reading.\\
\indent When looking at the table we note the following fluctuations: for observation no. 13 the value $q_1 = 12.2$ in instrument 1 remains about 40 percent below the average. At the same time instrument 2 registers a decrease of about 2 ions. Thus, there should indeed be a fluctuation of the radiation here, although the decrease by 5 ions indicated by instrument 1 may be partly falsified by reading errors. The fact that instrument 3 did not indicate a major decrease is due to the perhaps somewhat different behavior of the other radiation in height, which will be discussed later.\\
\indent An unquestionably real fluctuation of the radiation is to be noticed for observation no. 10 between 4:30 and 5:30 am. In all three instruments an increase in radiation of 2.8, 3.4 and 2.0 ions, respectively, is found simultaneously. However, this does not coincide in any case with when approaching the surface of the Earth to 70 m, as this would contradict all of my experiences made during other flights.\\
\indent Since the fluctuations were not accompanied by  meteorological changes, they can hardly be attributed to a change in the distribution of radioactive substances in the atmosphere.\\
\indent As the balloon continued to rise (obs. no. 11-14), in general a slight increase in radiation was again observed. With the exception of observation no. 13, which is influenced by the fluctuation of the radiation, an average of $q_1 = 17.6, q_2 = 10.5, q_3 = 20.8$ results at altitudes of 800-1600 m. The values are about the same order as those found before the ascent.\\

\begin{center}{3$^{\mathrm{rd}}$ Flight.}\end{center}

\indent The ascent took place on May 20, 1912 at 10:10 pm; rapidly we went via Korneuburg, Neuh\"ausl at the Thaya towards NNW. At morning clouds appeared. At 4 am Kuttenberg in B\"ohmen was reached. Because of lack of ballast we landed at 5:30 am descending from  a maximum height of 1200 m in Sadowa-Dohalice to the north of K\"onigsgr\"atz.\\
\indent The mean values of the radiation are:
\begin{center}
\tiny
\begin{tabular}{llll}
&Instrument 1&Instrument 2& Instrument 3\\
Before ascent in Vienna&$q_1=16,9$&$q_2=11,4$&$q_3=19,7$\\
During the night from&&&\\
10:30-2:30 am at
190 m above&&&\\
150-340 m altitude&$q_1=16,9$&$q_2=11,1$&$q_\mathrm{3red.}=18,2$\\
2:30-4:30 am at ca.&&\\
500 m relative altitude&$q_1=14,7$&$q_2=9,6$&$q_\mathrm{3red.}=17,6$\\
\end{tabular}
\end{center}

\indent The observations at 150-340 m altitude differ very little from those on the ground. Probably, the normal decrease of the $\gamma$-radiation is obscured by accidental increase of the residual radiation from the surface of the Earth. On the other hand, the decrease during the measurements at 500 m asl clearly shows that the ionization difference with respect to the values before ascent amounts to 2.2, 1.8, and 2.1 ions for the 3 instruments, respectively. Between 11:30 and 12:30 in the evening a radiation fluctuation, namely an increase by 2.8 or 2.0 and 1.0 ions  respectively, was found in all three instruments.\\

\begin{center}{4$^{\mathrm{th}}$ Flight.}\end{center}
\indent The ascent took place on June 3 at 9:45 in the evening. It was a 2200 cbm {\em [Translator's note: cubic meters]} balloon available, with which high  altitudes could be reached. Unfortunately, an approaching storm forced us to already land at 1:30 at night. The maximum height was 1900 m (absolute). Observations were made only from 10:30 to 12:30 at night at altitudes from 800 to 1100 m above the ground. The obtained mean values are $q_1 = 15.5, q_2 = 11.2, q_{3,\mathrm{red.}} = 21.8$, from which the values before the ascent, i.e, $q_1 =
15.5, q_2 = 11.7, q_3 = 21.3$ showed very little deviations.\\

\begin{center}{5$^{\mathrm{th}}$ Flight.}\end{center}
\indent This took place June 19 at 5 pm. Since I had to ascend on my own and thus had to take care of the guidance of the balloon, I only took one radiation detector with me. At 850 to 950 m relative altitude I found $q_2 = 9.8$ to 10.7 ions, while in the two hours prior to the ascent I observed $q_2 = 12.3$ to 14.5. The average at 900 m altitude is about 3 ions lower than at the ground.

From the results of the 4th and 5th flight we shall conclude that {\em an observable decrease of the radiation is likely to reach up at 1000 m above the ground, which however may be under certain circumstances - as during the 4th flight - obscured by accidental increase of the residual radiation.}\\

\begin{center}{6$^{\mathrm{th}}$ Flight.}\end{center}

\indent During this flight the radiation in the immediate vicinity of the Earth had to be examined again. I ascended on June 28 at 11:30 pm. The balloon was kept at heights from 280 to 360 meters for 5 hours. I only took the two thick-walled instruments with me. Their data was completely simultaneous throughout the night. Between 1:30 and 2:21 at night there was an increase in radiation of 1.9 and 1.6 ions, respectively, without a change in the altitude of the balloon or any change in the meteorological conditions. At 360 m above the ground the radiation decrease compared to the values on the ground was of  2.1 ions in instrument 1, 2.4 ions  in instrument 2. These results confirm the conclusions earlier  drawn.\\
\indent The following 7th flight was undertaken as a really high-altitude ascent.\\

\begin{center}{7$^{\mathrm{th}}$ Flight.}\end{center}

\begin{table*}[h!]
\begin{center}
\tiny
\begin{tabular}{c|c|c|c|c|c|c|cl|lc}
\multicolumn{5}{l}{Balloon: ``B\"ohmen'' (1680 cbm hydrogen)}&\multicolumn{6}{r}{Driver: Captain W. Hoffory.}\\
\multicolumn{5}{l}{Meteorolog. observer: E. Wolf.}&\multicolumn{6}{r}{Observer of atmospheric electricity: V. F. Hess.}\\\hline\hline
 &&\multicolumn{2}{c|}{Average altitude}&\multicolumn{5}{c|}{Observed radiation}&&Rel.\\
 \cline{5-9}
 No.&Time&\multicolumn{2}{c|}{}&Instrument 1&Instrument 2&\multicolumn{3}{c|}{Instrument 3}&Temp.&hum.\\
  \cline{3-4}
 &&absolute&relative&&&\multicolumn{3}{c|}{}&&percent\\ 
  \cline{5-9}
   &&m&m&$q_1$&$q_2$&$q_3$&red. $q_3$&&\\\hline\hline
  1&15$^\mathrm{h}$ 15 - 16$^\mathrm{h}$ 15 &156&0&17.3&12.9&-&-&\multirow{3}{*}{\huge\}}&-&-\\
     2&16$^\mathrm{h}$ 15 - 17$^\mathrm{h}$ 15 &156&0&15.9&11.0&18.4&18.4&&\multicolumn{2}{c}{1 $^1/_2$ days before}\\
      3&17$^\mathrm{h}$ 15 - 18$^\mathrm{h}$ 15 &156&0&15.8&11.2&17.5&19.5&&\multicolumn{2}{c}{ascent (in Vienna)}\\
 4&6$^\mathrm{h}$ 45 - 7$^\mathrm{h}$ 45 &1700&1400&15.8&14.4&21.1&20.0&&+6.4$^{\circ}$&60\\
 5&7$^\mathrm{h}$ 45 - 8$^\mathrm{h}$ 45 &2750&2500&17.3&12.3&22.5&19.8&&+1.4$^{\circ}$&41\\
 6&8$^\mathrm{h}$ 45 - 9$^\mathrm{h}$ 45 &3850&3600&19.8&16.5&21.8&17.9&&-6.8$^{\circ}$&64\\
7&9$^\mathrm{h}$ 45 - 10$^\mathrm{h}$ 45 &4800&4700&40.7&31.8&&&&-9.8$^{\circ}$&40\\
 &&(4400-5350)&&&&&-&-&-&-\\
 8&10$^\mathrm{h}$ 45 - 11$^\mathrm{h}$ 15 &4400&4200&28.1&22.7&-&-&&-&-\\
 9&11$^\mathrm{h}$ 15 - 11$^\mathrm{h}$ 45 &1300&1200&(9.7)&11.5&-&-&&-&-\\
 10&11$^\mathrm{h}$ 45 - 12$^\mathrm{h}$ 10 &250&150&11.9&10.7&-&-&&+16,0$^{\circ}$&68\\
 11&12$^\mathrm{h}$ 25 - 12$^\mathrm{h}$ 12 &140&0&15.0&11.6&-&-&&\multicolumn{2}{|l}{(after the landing in}\\
&&&&&&&&&\multicolumn{2}{|l}{Pieskow, Brandenburg)}
 \end{tabular}
\end{center}
\end{table*}

\indent We ascended at 6:12 am from Aussig on the Elbe. We flew over the Saxon border near Peterswalde, Struppen close to Pirna, Bischofswerda and Kottbus. In the region of the Schwielochsee the altitude of 5350 m was reached. At 12:15 at noon we landed at Pieskow, 50 km east of Berlin. Unfortunately, no observations could be made at the place of ascent before the flight. However, measurements were taken after landing under the unopened balloon to see if the balloon that had descended directly from 5000 m was covered with radioactive induction and had emitted radiation itself. As shown in the table (obs. no. 11), no trace of an increase in radiation was noticed under the landed balloon. The weather was not completely clear on this flight. A barometric depression approaching from the West became noticeable through the onset of clouds. But it should be explicitly noted that we were never in a cloud, not even near, since at the time when the cumulus clouds appeared as isolated balls across the horizon, we were already above 4000 m asl. Above us, when we were already at maximum altitude, was a much higher thin layer of clouds, whose lower limit could have been at least 6000 meters. The Sun was only shining through faintly.\\
\indent Let us first look at the results of the thick-walled instruments 1 and 2. At 1400 to 2500 m of average altitude the radiation was approximately as high as usually found on the ground. Then, however, an increase in radiation, clearly noticeable in both instruments, begins with increasing altitude. At 3600 m above the ground, the values are already 4-5 ions higher than on the ground.

\indent As far as the results of the thin-walled instrument 3 are concerned, it seems that the increase in radiation was already noticeable at even lower altitude.

However, given the uncertainty discussed previously, which involves the reduction of the values of this instrument to normal pressure\footnote{I did not apply a correction due to changes in the absolute temperature; for the flights no. 1-6 this change was never worth of consideration. However, even for the measurements just discussed, it has no consequence because the temperature in the measuring area of the instrument could decrease much less than the outside temperature measured by means of an aspiration thermometer because of the Sun's radiation.}, one may not consider this conclusion to be quite certain. Incidentally, the increase is also qualitatively noticeable in the unreduced values $q_3$. The readings on instrument 3 have an unintentional end at 10:45, since the ionization cylinder loosened due to an unskilful manipulation just before the reading at the maximum height, and the electrometer discharged itself by contact with the {central pin}.

{\em In the two $\gamma$-ray detectors the values at the maximum height are 20 to 24 ions higher than on the ground. During the descent, still  very high values of $q_1 = 28.1$ and $q_2 = 22.7$ were found at 4400 m of average absolute altitude.} These values exceed the normal values as well by 12 or 11, respectively. In the subsequent very rapid fall (2 m per second) a very low value of 9.7 was measured at 1200 m of average height in instrument 1, while in instrument 2 the normal value of 11.5 was registered. I think it is possible that in instrument 1, which has with very thick fibers, sometimes a certain stiffness of the fibers can be disturbing.\\
\indent The results of both instruments obtained under the still-filled balloon after landing  are, as already noted above, quite normal.\\
\indent In order to gain an overview of the change in penetrating radiation with altitude as described by the mean values, in the following table I have compiled all the 88 radiation values measured by me in the balloon according to altitude levels. Since here, for each altitude range, mean values of several individual values are calculated, which were obtained under different circumstances and may be influenced by the temporal variations already mentioned, one cannot expect to get already a very accurate picture of the course of the radiation with increasing altitude. The numbers enclosed in parentheses next to the radiation values represent the number of observations from which the mean value was determined.

\begin{center}
\tiny
\begin{tabular}{c|c|c|c|c}
\hline\hline
\multicolumn{5}{c}{Table of the mean values}\\
\hline\hline
Average altitude above &\multicolumn{4}{l}{Observed radiation in ions per cm$^3$ and second.}\\
\cline{2-5}
the ground&Instrument 1&Instrument 2&Instrument 3&\\
\cline{2-5}
m&$Q_1$&$Q_2$&$Q_3$&$Q_3$\\
&&&(reduced)&(not reduced)\\
\hline\hline
0&16.3 (18)&19.6 (9)&11.8 (20)& 19.7 (9)\\
up to 200&15.4 (13)&11.1 (12)&19.1 (8)& 18.5 (8)\\
200-500&15.5 (6)&10.4 (6)&18.8 (5)& 17.7 (5)\\
500-1000&15.6 (3)&10.3 (4)&20.8 (2)& 18.5 (2)\\
1000-2000&15.9 (7)&12.1 (8)&22.2 (4)& 18.7 (4)\\
2000-3000&17.3 (1)&13.3 (1)&31.2 (1)& 22.5 (1)\\
3000-4000&19.8 (1)&16.5 (1)&35.2 (1)& 21.8 (1)\\
4000-5200&34.4 (2)&27.2 (2)&-& -\\
  \end{tabular}
\end{center}

 We see from the table that immediately above the ground the total radiation decreases a little --  0.8 to 1.4 ions on average. Since, however, a decrease of up to 3 ions has sometimes been found during the individual flights, and in the case of many measurements one of over 2 ions, we will quote about 3 ions as the maximum value of the decrease. This decrease extends up to  1000 m above the ground. As mentioned, it is apparently due to the absorption of the $\gamma$ rays emanating from the surface of the Earth. From this we conclude that {\em the $\gamma$-radiation of the Earth's surface and of the uppermost soil layers causes a ionization in zinc vessels of about 3 ions per cm$^3$ per second.}
 
 At altitudes up to 2000 m there is already a noticeable increase in radiation again. The increase reaches the value of 4 ions from 3000 to 4000 m, and even 16 to 18 ions from 4000 to 5200 m in both instruments. For the thin-walled instrument 3 the decrease, when the values are reduced to normal pressure, is reversed sooner and more strongly.\\
\indent Where does this increase of penetrating radiation with height, which was simultaneously and repeatedly observed with the three instruments, originate from?\\
\indent Assuming that only the known radioactive substances in the Earth's crust and in the atmosphere emit radiation with the characteristics of $\gamma$-radiation and produce ionization in closed vessels, there are great difficulties in finding an explanation.\\
\indent According to the direct determination of the absorption coefficient of  $\gamma$ rays in the air by myself\footnote{Vienna, Session Rep. 120, 1205-1212, 1911.}  and Chadwick\footnote{Le Radium 9, 200 -202, 1912.}, the absorption of the radiation emanating from the Earth's surface  takes place quite rapidly, so that at 500 m above the ground hardly more than 10 percent of the radiation can remain. As mentioned, I have also been able to demonstrate this decrease experimentally in the balloon, although it has been found that the radioactive substances of the Earth's surface do not play such a predominant role in the total radiation, as some authors believed. Their fraction was determined with 3 ions per cubic centimeter per second.\\
\indent The decay products of the emanations are still left as possible $\gamma$-radiation  ionizers  at height. Because of their short lifetime, the emanations of thorium and actinium as well as their decay products cannot reach high altitudes. Only the radium emanation with a half life of almost 4 days will be able to be uplifted by rising air currents to high altitudes. In general, however, the concentration of the emanation and thus also the content of radium G in the air will soon decrease with altitude. An increase of the radiation with height could occur only with a random accumulation of RaC of purely local character: it would for instance be possible that such accumulations occur in stable layers with temperature inversion, also in cumulus clouds or in mist, since it is known that the RaC atoms frequently act as condensation nuclei. Apparently, the steady increase in penetrating radiation with altitude,  shown by my observations, cannot be explained in this way. In addition, in flight no. 2 and 6 where the balloon flew for hours on a stable layer with temperature inversion close to the Earth, I did not observe any increase in radiation although the content of RaC in air must be higher close to the ground. At a height of 5000 m the RaC content will not be sufficient at all to cause as much radiation increase as I found.\\
\indent The fluctuations of the radiation found by Pacini\footnote{Op. cit.} and Gockel\footnote{Jahrb. d. Rad. u. Elektron 9, 1 - 15, 1912.}  on the sea and on shore, which I also found often in the balloon, cause some difficulties regarding the explanation of he penetrating radiation on the sole basis of the radioactive theory. I have repeatedly observed such fluctuations in the middle of the night, with perfectly calm atmosphere. In the absence of any meteorological change there is no reason to attribute it to changes in the distribution of the radioactive substances of the atmosphere.

{\em The results of the present observations seem most likely to be explained by the assumption that radiation of very high penetrating power enters from above into our atmosphere, and even in its lowest layers  causes part of the ionization observed in closed vessels. The intensity of this radiation appears to be subject to temporal fluctuations, which are still recognisable at one-hour reading intervals.} Since I found no decrease in the radiation either at night or during a solar eclipse, one can hardly consider the Sun as the cause of this hypothetical radiation, at least as long as one only imagines a direct $\gamma$-radiation with rectilinear propagation.\\
\indent It is not so surprising that the increase of the radiation becomes noticeable only beyond 3000 m: in the first 1000 m the decrease of the $\gamma$-radiation at the Earth's surface predominates, and then the decrease of the induction content joins in, which certainly has an impact up to more than 3000 m. In any case the absorption of the radiation coming from above follows an exponential curve; therefore, the increase of radiation from the bottom to the top will be steeper at high altitudes.\\
\indent Finally, let us consider the experiments by Wright, Simpson, Mc. Lennan, Wulf that have among other things led to the opinion that the penetrating radiation at the Earth's surface almost exclusively arises from the radioactive substances from the ground and not from the atmosphere. The same authors agreed that over water or ice already at a short distance from the land, the radiation is 4-6 ions lower. Hence, they found a larger difference than I  between the values on the ground and those at a few hundred meters altitude (2- 3 ions). My results are in perfect agreement with one of the observations by Prof. Wulf, who found on the top of the Eiffel Tour in Paris a radiation of 2.3 ions lower than on the ground. It is known that $\gamma$-radiation also excites secondary $\beta$ and $\gamma$ rays when hitting matter\footnote{A. Brommer, Vienna, session report July 1912.}. However, the secondary radiation excited on solid bodies is stronger than the secondary radiation generated on a water surface. The radiation coming from above will certainly be able to excite secondary radiation at the Earth's surface, but above land the generated secondary radiation will be larger than above water. This may clarify the difference in the estimation of the Earth's radiation from the balloon experiments and those above water.\\
\indent The investigations so far have shown that the penetrating radiation observed in closed vessels is of very complex origin. Part of the radiation comes from the radioactive substances on the Earth's surface and in the uppermost soil layers and  fluctuates relatively little. A second component, which is influenced by meteorological factors,  arises from the radioactive substances of the atmosphere, essentially from RaC. My balloon observations seem to indicate that there is still a third component of the total radiation, which increases in height and also shows remarkable intensity fluctuations on the ground. Further research will have to pay the largest attention especially to these fluctuations.\\
\indent I would like to express my heartfelt thanks to the imperial-royal Austrian Aeroclub and the management of the German Aviation Association in B\"ohmen for the sympathetic support of my work.

\newpage

\title{\bf{Comments to ``On the Observations of the Penetrating\\Radiation during Seven Balloon Flights''}}
\author{\bf{Article by V.F. Hess: Physikalische Zeitschrift  13 (1912) 1084}}
\vskip 3 mm
\date{Translated and commented by  A. De Angelis and C. Arcaro b. Schultz\\INFN and INAF Padova, Italy}

\maketitle

\newcommand{\BY}[1]{{#1},}
\newcommand{\IN}[4]{{#1} \textbf{#2} (#3) #4}

\begin{abstract}
At the beginning of the twentieth century,  the
Austrian (later naturalized American) Victor Hess among others
developed a brilliant line of research \cite{noi2,iospringer}, leading to the final
determination of the extraterrestrial origin of part of the atmospheric radiation. Before his
work, the origin of the radiation  today called ``cosmic rays''  was
strongly debated, as many scientists thought that these particles
came integrally from the crust of the Earth. There was however an active and rich research on the topic.

Victor (at that time Viktor) Hess measured in 1912 the rate of
discharge of an electroscope that flew aboard an atmospheric balloon.
Since the discharge rate increased as the balloon flew at higher
altitude, he concluded   that the origin of part of the natural radiation could not be
terrestrial. For this discovery, Hess was awarded the Nobel Prize in
1936, and his experiment became legendary.

The fundamental article published by Victor Hess in 1912, reporting a significant increase of the radiation as altitude increases, is translated and commented here. The decisive measurement was performed in  the last of the seven flights on aerostatic balloon described.

\end{abstract}



Already in 1785  Coulomb found \cite{Cou1785} that electroscopes can spontaneously discharge by the action of the air and not by defective insulation. After dedicated studies by Faraday around 1835 \cite{Far1835}, Crookes observed in 1879 \cite{Cro1879} that the speed of discharge of an electroscope decreased when pressure was reduced. The explanation of the phenomenon came in the beginning of the 20th century and paved the way to one of mankind's revolutionary scientific discoveries: cosmic rays. 


Spontaneous radioactivity was discovered in 1896 by Becquerel \cite{Beq1896}. A few years later, Marie and Pierre Curie \cite{Cur1898} discovered that the elements Polonium and Radium suffered transmutations generating radioactivity: such transmutation processes were then called ``radioactive decays''. In the presence of a radioactive material, a charged electroscope promptly discharges. It was concluded that some elements were able to emit charged particles, that in turn caused the discharge of the electroscopes.  The discharge rate was then used to gauge the level of radioactivity. 

Following the discovery of radioactivity, a new era of research in discharge physics was then opened, this period being strongly influenced by the discoveries of the electron and of positive ions. During the first decade of the 20th century results on the study of ionization phenomena came from several researchers in Europe and in the United States. 

Around 1900, Wilson \cite{Wil1901} and Elster and Geitel \cite{ElG1900} improved the technique for a careful insulation of electroscopes in a closed vessel, thus improving the sensitivity of the electroscope itself. As a result, they could make quantitative measurements of the rate of spontaneous discharge. They concluded that such a discharge was due to ionizing agents coming from outside the vessel. The obvious questions concerned the nature of such radiation, and whether it was of terrestrial or extra-terrestrial origin. The simplest hypothesis was that its origin was related to radioactive materials, hence terrestrial origin was a commonplace assumption. An experimental proof, however, seemed hard to achieve. Wilson \cite{Wil1901} tentatively made the visionary suggestion that the origin of such ionization could be an extremely penetrating extra-terrestrial radiation. However, his investigations in tunnels with solid rock overhead showed no reduction in ionization \cite{Wil1901} and could therefore not support an extra-terrestrial origin. An extra-terrestrial origin, though now and then discussed (e.g. by Richardson \cite{Ric1906}), was dropped for the following years.

In 1903 Rutherford and Cook \cite{RuC1903} and also McLennan and Burton  \cite{MLB1903} showed that the ionization was significantly reduced when the closed vessel was surrounded by shields of metal kept free from radioactive impurity. This showed that part of the radiation came from outside. Later investigations showed that the ionization in a closed vessel was due to a ``penetrating radiation'' partly from the walls of the vessel and partly from outside. 

The situation in 1909 is well summarized by Kurz \cite{kurz}  and by Cline \cite{Cline}.  The spontaneous discharge was consistent with the hypothesis that even in insulated environments a background radiation did exist. In the 1909 review by Kurz three possible sources for the penetrating radiation are discussed: an extra-terrestrial radiation possibly from the Sun \cite{Ric1906}, radioactivity from the crust of the Earth, and radioactivity in the atmosphere. Kurz concludes from the ionization measurements in the lower part of the atmosphere that an extra-terrestrial radiation is unlikely.
It was generally assumed that large part of the radiation came from radioactive material in the crust. Calculations were made of how the radiation should decrease with height (see, {\em{e.g.,}} Eve \cite{eve}) and measurements were performed.

\begin{figure}
\begin{center}
\resizebox{0.5\columnwidth}{!}{\includegraphics{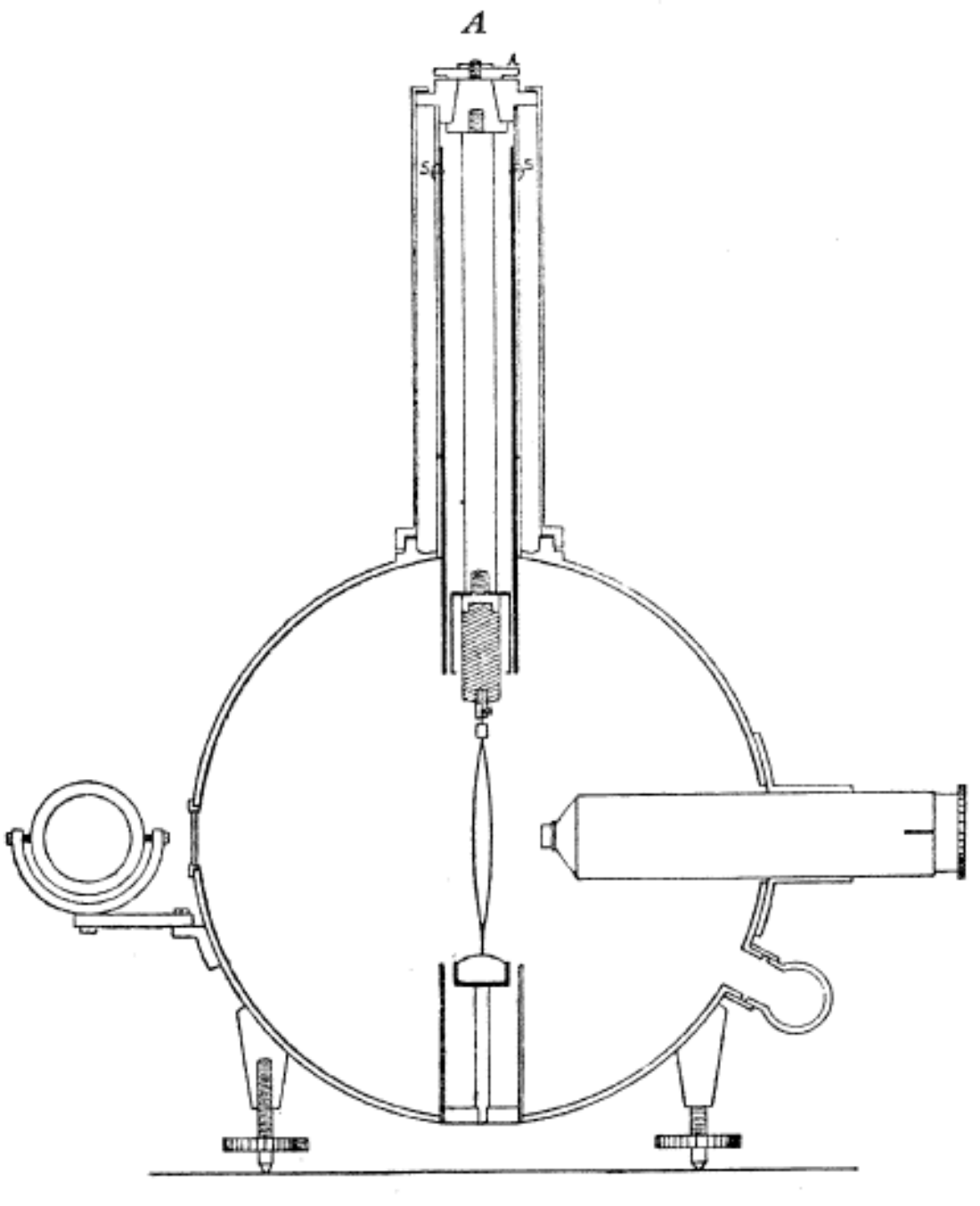} }
\end{center}
\caption{The Wulf electroscope. The 17 cm diameter cylinder with depth 13 cm was made
of Zinc. To the right is the microscope that measured the distance between the two silicon
glass wires illuminated using the mirror to the left. The air was kept dry using Sodium in the small container below the microscope. According to Wulf \cite{Wul1910}, with 1.6 ion pairs
per second produced, the tension was reduced by 1~V, the sensitivity of the instrument,
as measured by the decrease of the inter-wire distance.}
\label{fig:1}       
\end{figure}

Father Theodor Wulf, a German scientist and a Jesuit priest serving in the Netherlands and then in Roma,  had the idea to check the variation of radioactivity with height to test its origin. First he improved the electroscope (Fig. \ref{fig:1}) replacing the two leaves   by two metalized silicon glass wires (called ``fibers'' in the translation of Hess' article), with a tension spring  in between: with this improvement the instrument became more easily transportable. Then in 1909 \cite{Wul1910} he measured the rate of ionization at the top of the Eiffel Tower in Paris (300 m above ground). Supporting the hypothesis of the terrestrial origin of most of the radiation, he expected to find at the top  much  a smaller ionization than on the ground. The rate of ionization showed, however, too small a decrease to confirm the hypothesis. He concluded that, in comparison with the values on the ground, the intensity of radiation ``decreases at nearly 300 m [altitude] not even to half of its ground value''; while with the assumption that radiation emerges from the ground there would remain at the top of the tower ``just a few percent of the ground radiation'' \cite{Wul1910}. Wulf's observations were of great value, because he could take data at different hours of the day and for many days at the same place. For a long time, Wulf's data were considered as the most reliable source of information on the altitude effect in the penetrating radiation. However Wulf concluded that the most likely explanation of his puzzling result was still emission from the soil.

Other measurements with similar results were also made (Bergwitz \cite{Bergwitz}, McLennan and Macallum \cite{McLennan}, Gockel \cite{Goc1909}). The general interpretation of the outcome was that radioactivity was mostly coming from the Earth's crust.


The conclusion that radioactivity was mostly coming from the Earth's crust was questioned by many, in particular by t Domenico Pacini, who compared the rate of ionization on mountains, over a lake, and over the sea \cite{Pac1909,Pac1910}; in 1911, he made important experiments by immersing an electroscope deep in the sea \cite{Pac1912}. 

In a first period Pacini made several measurements to establish the variations of the electroscope's discharge rate as a function of the environment. First he placed the electroscope on the ground and on a sea a few kilometers off the coast. Pacini made his measurements over the sea in the Gulf of Genova, on an Italian Navy ship, the {\em cacciatorpediniere} (destroyer)  ``Fulmine"  from the Accademia Navale di Livorno; the results were comparable. A summary of these results indicate, according to Pacini's conclusions, that ``in the hypothesis that the origin of penetrating radiations is in the soil, since one must admit that they are emitted at an almost constant rate (at least when the soil is not covered by remaining precipitations), it is not possible to explain the results obtained'' \cite{Pac1909}. Pacini's conclusion, confirmed by Gockel \cite{Goc1909}, was the first where it was established that the results of many experiments on radiation could not be explained by radioactivity in the Earth's crust.

Pacini continued the investigations of radiation and developed in 1911 a new experimental technique for the measurement of radioactivity a
few meters underwater;
 he found a significant decrease
by 20\% in the discharge rate when the electroscope was placed three meters underwater in the sea in front of the Naval Academy of Livorno (and later in the Lake of Bracciano), consistent with absorption by water of a radiation coming from outside, and he cincluded  \cite{Pac1912}. 
that: ``[It] appears from the results of the work described in this Note that a sizeable cause of ionization exists in the atmosphere, originating from penetrating radiation, independent of the direct action of radioactive substances in the soil." 
Pacini however could not firmly disprove a possible atmospheric origin of the background radiation.

The need for balloon experiments \cite{libropd} became evident to clarify Wulf's observations on the effect of altitude (at that time and since 1885, balloon experiments were anyway widely used for studies of the atmospheric electricity). The first balloon flight with the purpose of studying the properties of penetrating radiation was arranged in Switzerland in December 1909 with a balloon called Gotthard from the Swiss aeroclub. Alfred Gockel, professor at the University of Fribourg, ascending up to 4500 m above sea level (a.s.l.) during three successive flights, found \cite{Goc1910,Goc1911} that the ionization did not decrease with height as expected on the hypothesis of a terrestrial origin. Gockel confirmed the conclusion of Pacini in \cite{Pac1909}, quoting him
correctly, and concluded ``that a non-negligible part of the penetrating radiation is independent of the direct action of the radioactive substances in the uppermost layers of the Earth'' \cite{Goc1911}. In a 1909 balloon ascent Bergwitz had found \cite{Bergwitz} that the ionization at 1300 m altitude had decreased to about 24\% of the value at ground, consistent with expectations if the radiation came from the EarthÕs surface. However, Bergwitz's results were questioned because the electrometer was damaged during the flight (see, e.g., \cite{Goc1911}).

In spite of Pacini's conclusions, and of Wulf's and Gockel's puzzling results on the dependence of radioactivity on altitude, physicists were however reluctant to give up the hypothesis of a terrestrial origin. The situation was cleared up thanks to a long series of balloon flights by the Austrian physicist Victor Hess, who established the extra-terrestrial origin of at least part of the radiation causing the observed ionization. 

Hess was born in 1883 in Steiermark, Austria, and he took his doctor's degree in 1910 in Graz. After graduation he was assistant under professor Meyer at the Institute of Radium Research of the Viennese Academy of Sciences, where he performed most of his work on cosmic rays, and in 1919 he became Professor of Experimental Physics at the Graz University. Hess was on leave of absence from 1921 to 1923 and worked in the United States, where he took a post as Director of the Research Laboratory of the United States Radium Corporation, at Orange (New Jersey). In 1923 he returned to Graz and in 1931 he moved to Innsbruck as professor. In 1936 Hess was awarded the Nobel Prize in physics for the discovery of cosmic rays. After moving to United States of America (US) in 1938 as professor at Fordham University, Hess became an American citizen in 1944, and lived in New York until his death in 1964.

Hess, at that time working in Wien and in Graz, started his experiments by studying Wulf's results, and knowing the detailed predictions by Eve \cite{Eve1911} on the coefficients of absorption for radioactivity in the atmosphere. Eve wrote that, if one assumed a uniform distribution of RaC on the surface and in the uppermost layer of the Earth, ``an elevation to 100 m should reduce the [radiation] effect to 36 percent of the ground value''. There was  such a serious discrepancy with Wulf's results that its resolution appeared to be of the highest importance for the radioactive theory of atmospheric electricity \cite{Hes1912}. Since in the interpretation of Wulf's and Gockel's results the absorption length of the radiation (at that time identified mostly with gamma radiation) in air entered crucially, Hess decided first to improve the experimental accuracy of the Eve's result by direct measurements of the absorption of gamma rays in air \cite{Hes1911}. He used  probes of about 1 g RaCl$_2$ at distances up to 90 m, and obtained an absorption coefficient consistent with Eve. Hence the contradiction of Wulf's results remained; Hess concluded that ``a clarification can only be expected from further measurements of the penetrating radiation in balloon ascents'' \cite{Hes1911}.

\begin{figure}
\begin{center}
\includegraphics[width=0.4\columnwidth]{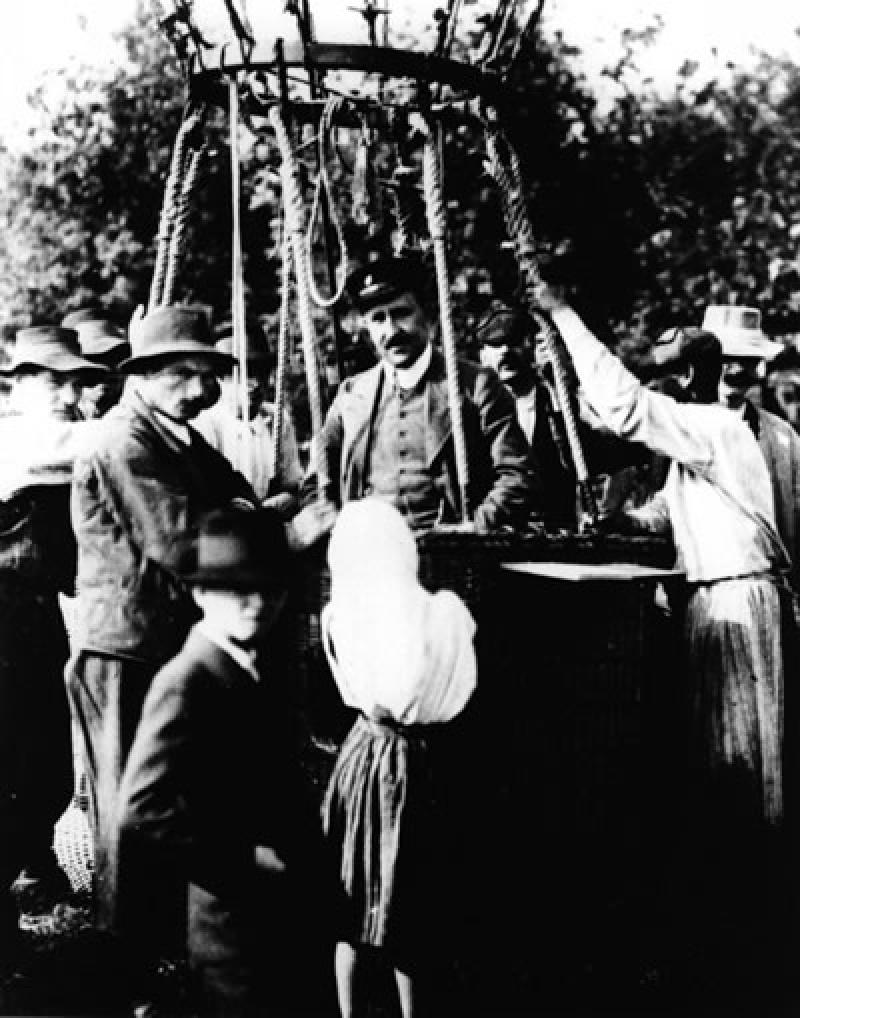}\includegraphics[width=0.53\columnwidth]{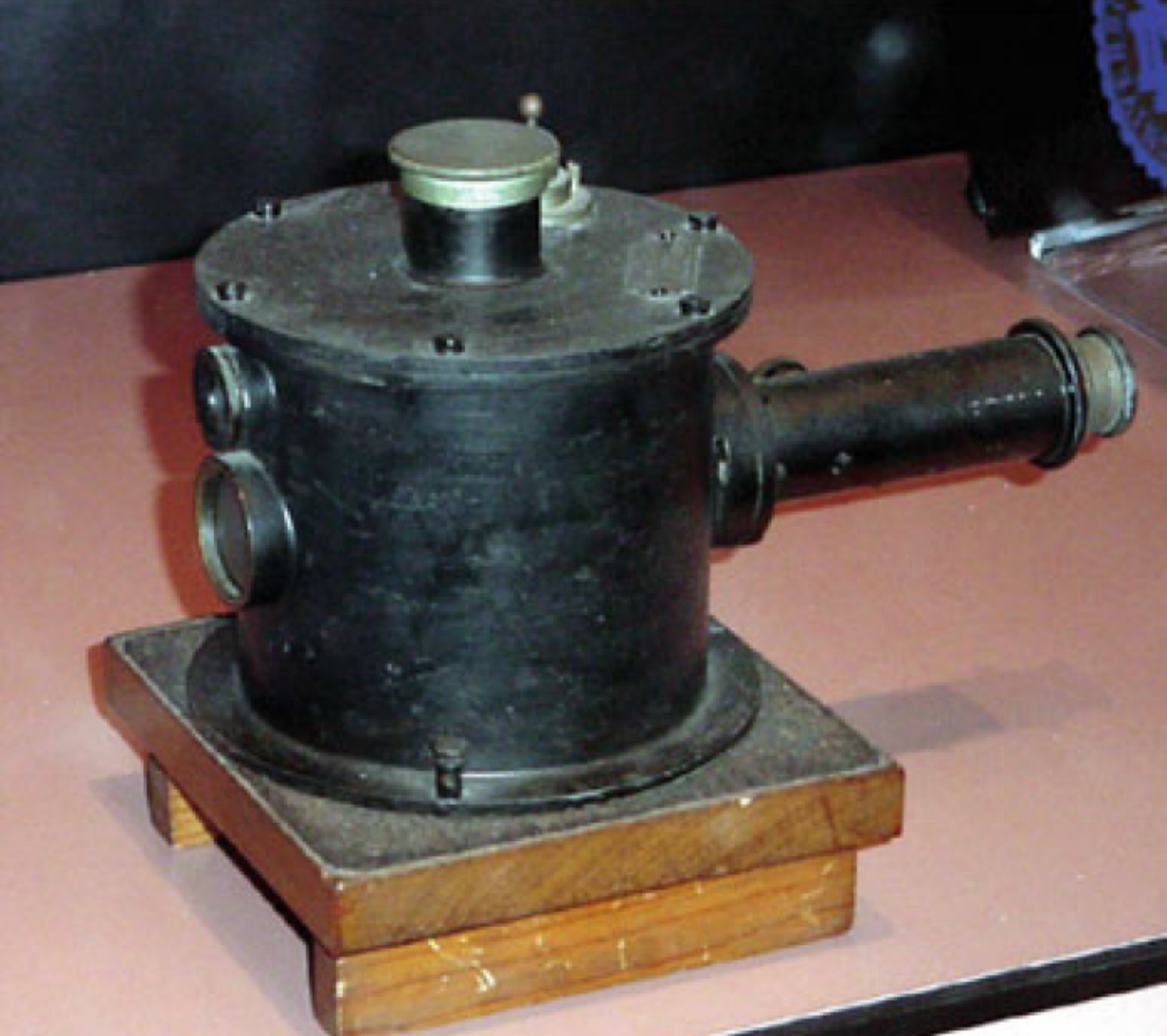} 
\end{center}
\caption{Left: Historical photograph of Hess' balloon flight. Right: one of the electrometers used by Hess during his flight. (Smithsonian National
Air and Science Museum, Washington, DC; photo by P. Carlson).}
\label{fig:balloon}       
\end{figure}

\begin{figure}
\begin{center}
\includegraphics[width=0.3\columnwidth]{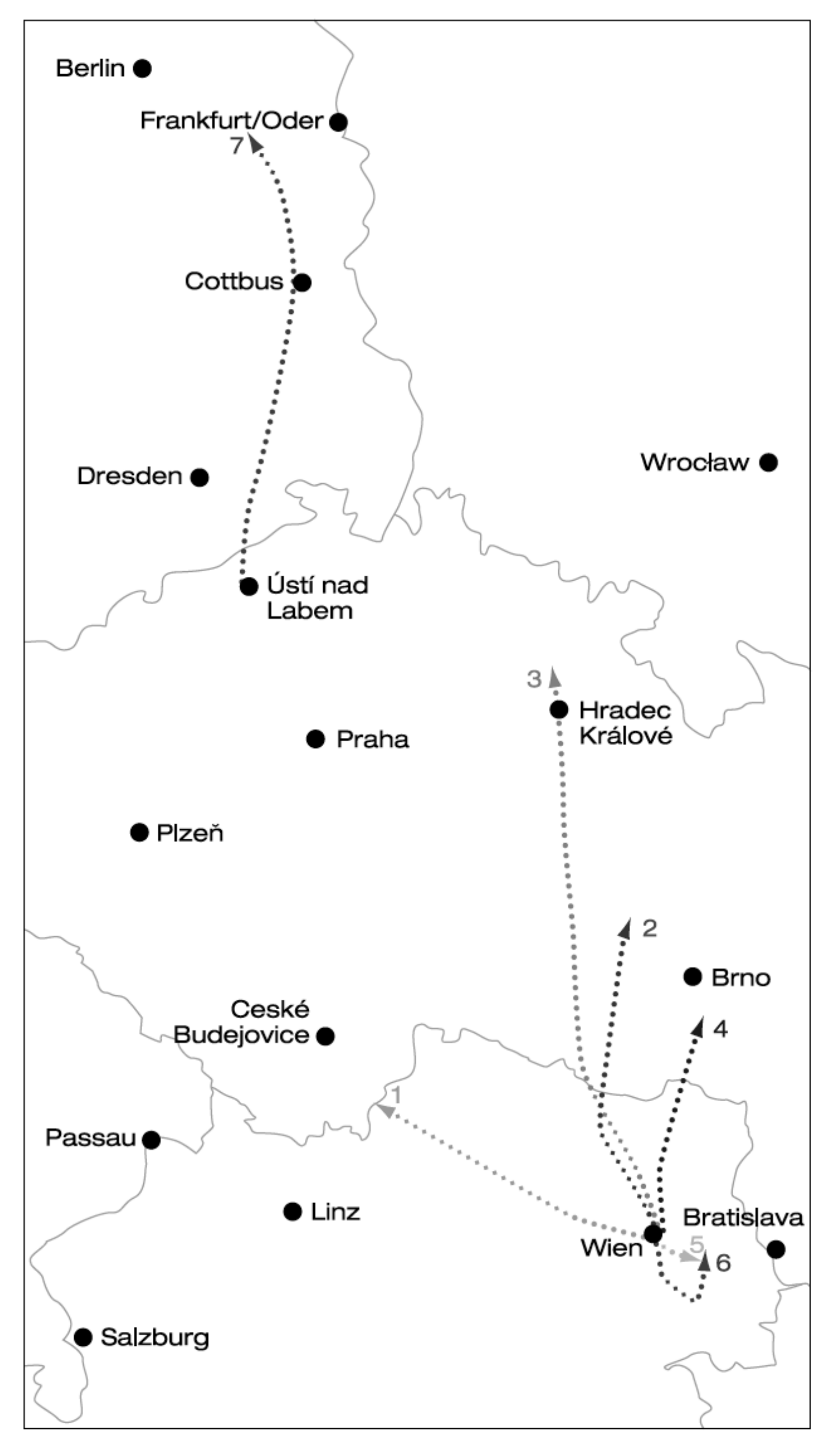} 
\includegraphics[width=0.68\columnwidth]{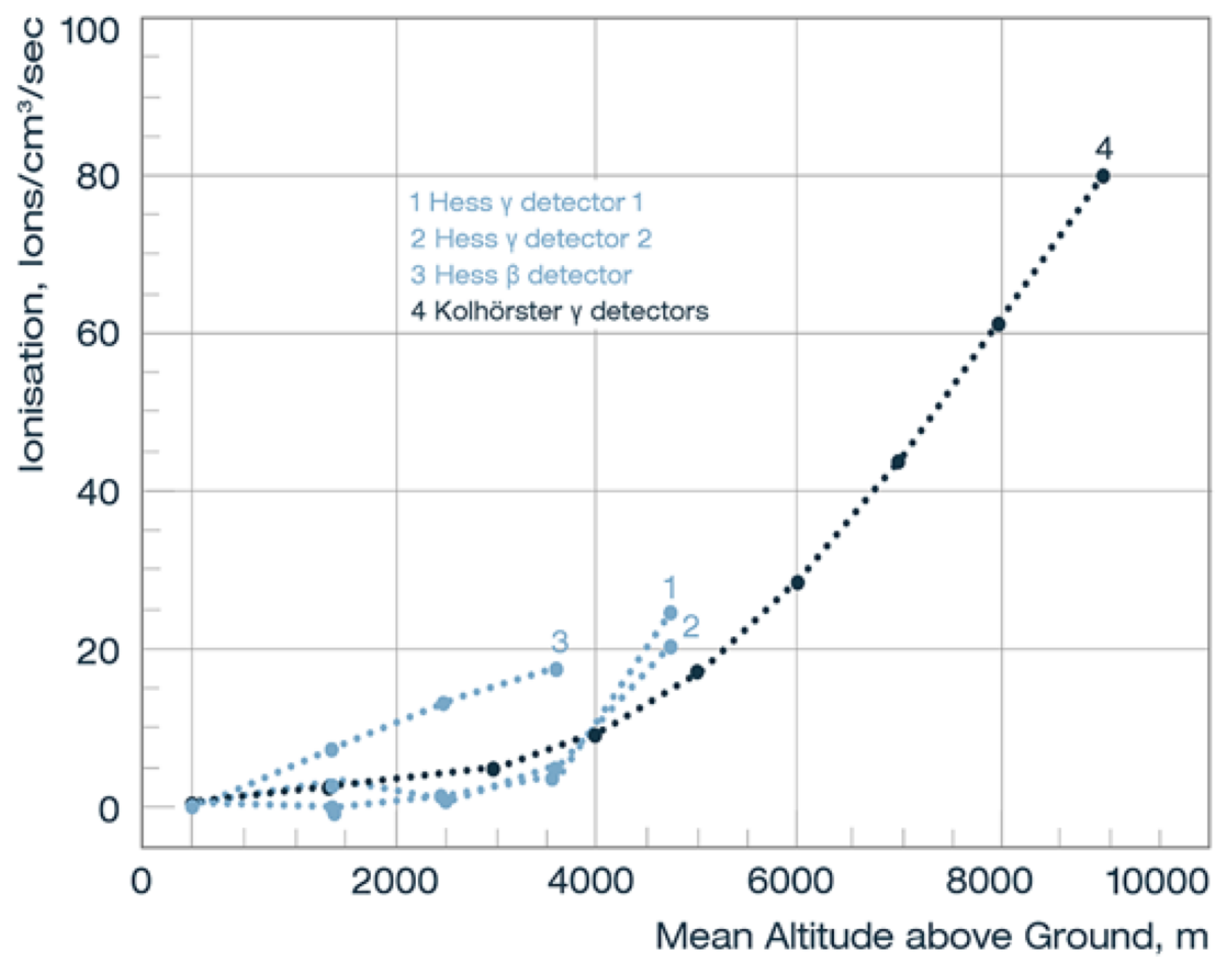} 
\end{center}
\caption{Left: The seven
balloon flights by Hess in
1912. Right: The ionization
measured in the 7th
high altitude flight of Hess
(1912) and in the flight of
Kolh\"rster in 1914 (average of his
two  detectors). The ionization
measured at the surface of the
Earth has been subtracted. Adapted from \cite{walter}.}
\label{fig:increase}       
\end{figure}

Hess continued his studies with balloon observations (Fig. \ref{fig:balloon}). 
From April 1912 to August 1912 Hess had the opportunity to fly seven times (Fig. \ref{fig:increase}, left) with three different instruments (enclosed in boxes with metal walls of
different thicknesses in order  to disentangle the effect of beta radiation). In the final flight, on August 7, 1912, he reached 5200 m. His results clearly showed that the ionization  increased considerably with height (Fig. \ref{fig:increase}, right).  (i) Immediately above ground the total radiation decreases a little. (ii) At altitudes of 1000 to 2000 m there occurs again a noticeable growth of penetrating radiation. (iii) The increase reaches, at altitudes of 3000 to 4000 m, already 50\% of the total radiation observed on the ground. (iv) At 4000 to 5200 m the radiation is stronger [more than 100\%] than on the ground \cite{Hes1912}. Note that the detector optimized for  $\beta$ radiation measurement broke during the experiment.


Hess concluded that the increase of the ionization with height must be due to a radiation coming from above, and he thought that this radiation was of extra-terrestrial origin. He also excluded the Sun as the direct source of this hypothetical penetrating radiation because of no day-night variation. Hess finally published a summary of his results in Physikalische Zeitschrift in 1913 \cite{Hes1913}, a paper which reached the wide public.

The results by Hess were later confirmed by Kolh\"orster \cite{Kol1914} in a number of flights up to 9200 m.  An increase of the ionization up to ten times that at sea level was found. The absorption coefficient of the radiation was estimated to be 10$^{-5}$ per cm of air at NTP. This value caused great surprise as it was 8 times smaller than the absorption coefficient of air for gamma rays as known at the time.


Hess is today remembered as the discoverer of cosmic rays for which he was awarded the 1936 Nobel Prize in physics, nominated by Compton. In his nomination Compton had written: ``The time has now arrived, it seems to me, when we can say that the so-called cosmic rays definitely have their origin at such remote distances from the Earth that they may properly be called cosmic, and that the use of the rays has by now led to results of such importance that they may be considered a discovery of the first magnitude. [...] It is, I believe, correct to say that Hess was the first to establish the increase of the ionization observed in electroscopes with increasing altitude; and he was certainly the first to ascribe with confidence this increased ionization to radiation coming from outside the Earth''. Why so late a recognition? Compton writes: ``Before it was appropriate to award the Nobel Prize for the discovery of these rays, it was necessary to await more positive evidence regarding their unique characteristics and importance in various fields of physics" \cite{noi2}.   The Nobel prize to Hess was shared with
C.D. Anderson for the discovery of the positron.

Hess' discovery was based on contributions of many  scientists; the contributions by Pacini, Wulf, Gockel and Eve were correctly cited in the final report by the Nobel prize Committee to the Royal Academy of Sweden  \cite{noi2}.

{\em Note - To our knowledge this is the first complete translation of Hess' paper. A partial translation was published in \cite{hillas}.}
%







\begin{thebibliography}{999}



\bibitem{noi2} \BY{P. Carlson and A. De Angelis} {\em Nationalism and internationalism in science: the case of the discovery of cosmic rays,}  
\IN{Eur. Phys. Eur. Phys. J. H} {35}{2010} {309} (arXiv:1012.5068 [physics.hist-ph]) 


\bibitem{iospringer} A. De Angelis,  {\em L'enigma dei raggi cosmici,} Springer 2010, DOI: 10.1007/978-88-470-2047-4, ISBN: 9788847020474
%
\bibitem{Cou1785} \BY{C. de Coulomb} {\em M\'em. Acad. des Sciences} (Paris, 1875) p. 612
\bibitem{Far1835} \BY{M. Faraday} {\em Researches in Electricity} (London, 1844)
\bibitem{Cro1879} \BY{W. Crookes} {\em On Electrical Insulation in High Vacua}, \IN{Proc. Roy. Soc.  London} {28} {1879} {347}
%
\bibitem{Beq1896} \BY{H. Becquerel}  {\em Sur les radiations \'emises par phosphorescence (On the radiation emitted by phosphorescence),} 
\IN{Comptes Rendus de l'Acad. des Sciences}{122}{1896} {420}
\bibitem{Cur1898} \BY{P. Curie, Mme P. Curie, and G. B\'emont}  {\em Sur une nouvelle substance fortement radio-active, contenue dans la pechblende (On a new, strongly radioactive substance contained in pitchblende),} 
\IN{Comptes Rendus de l'Acad. des Sciences}{127}{1898} {1215}
\bibitem{Wil1901} \BY{C.T.R. Wilson} {\em On the ionization of Atmospheric Air,} \IN{Proc. Roy. Soc.  London}{68}{1901}{151}
%
%
%
%
%
%
\bibitem{ElG1900} \BY{J. Elster and H.F. Geitel} \IN{Ann. d. Phys.}{2}{1900}{425}
\bibitem{Ric1906} \BY{I.G. Richardson} \IN{Nature} {73}{1906} {607}
\bibitem{RuC1903} \BY{E. Rutherford and H.L. Cook} \IN{Phys. Rev.}{16}{1903}{183}
\bibitem{MLB1903} \BY{F.C. McLennan and F. Burton} \IN{Phys. Rev.}{16}{1903}{184}
%
\bibitem{kurz} \BY{K. Kurz} \IN{Phys. Zeit.}{10}{1909} {834}
\bibitem{Cline}  \BY{G.A. Cline}  \IN{Phys. Rev.} {30} {1910} {35}
%
\bibitem{eve} \BY{A.S. Eve} \IN{Philos. Mag.} {13}{1907}{248}
%
\bibitem{Wul1910} \BY{Th. Wulf} \IN{Phys. Zeit.} {1}{1909}{152}
%
\bibitem{Bergwitz} \BY{K. Bergwitz} Hab. Paper (Braunschweig 1910)
\bibitem{McLennan}\BY{J.C. McLennan and E.N. Macallum} \IN{Phil. Mag.} {VI/22} {1911} {639}
%
\bibitem{Goc1909}   \BY{A. Gockel} \IN{Phys. Zeit.} {10}{1909} {845}
%
\bibitem{Pac1909} \BY{D. Pacini} {\em Sulle radiazioni penetranti,} \IN{Rend. Acc. Lincei} {18}{1909} {123} 
\bibitem{Pac1910} \BY{D. Pacini} {\em La radiazione penetrante sul mare,} \IN{Ann. Uff. Centr. Meteor.} {XXXII, parte I}{1910}{};  \IN{Le Radium} {VIII}{1911} {307} (translated and commented by M. De Maria and A. De Angelis in arXiv:1101.3015 [physics.hist-ph])
\bibitem{Pac1912} \BY{D. Pacini} {\em La radiazione penetrante alla superficie ed in seno alle acque,} 
\IN{Nuovo Cim.} {VI/3} {1912} {93} (translated and commented by A. De Angelis in arXiv:1002.1810 [physics.hist-ph]) 
%
\bibitem{libropd}  \BY{A. De Angelis, M. Pimenta} {\em Introduction to particle and astroparticle physics (multimessenger astronomy and its particle physics foundations)}, Springer Nature, Heidelberg 2018,
DOI: 10.1007/978-3-319-78181-5, ISBN: 978-3-319-78180-8
%
\bibitem{Goc1910} \BY{A. Gockel} \IN{Phys. Zeit.} {11}{1910} {280}
%
\bibitem{Goc1911}   \BY{A. Gockel} \IN{Phys. Zeit.} {12}{1911} {595}
%
\bibitem{Eve1911} \BY{A.S. Eve} \IN{Philos. Mag.} {21} {1911}{27}
%
%
%
%
\bibitem{Hes1912} \BY{V. Hess} \IN{Phys. Zeit.} {13}{1912} {1084}
\bibitem{Hes1911} \BY{V. Hess} \IN{Phys. Zeit.} {12}{1911} {998} 

\bibitem{Hes1913} \BY{V. Hess} \IN{Phys. Zeit.} {14}{1913} {612}
\bibitem{walter} M. Walter, in B. Falkenburg and W. Rhode, {\em From Ultra Rays to Astroparticles}, Springer, Heidelberg 2012
%
\bibitem{Kol1914} \BY{W. Kolh\"orster} \IN{Phys. Zeit.} {14} {1913}{1153}; \IN{Ber. deutsch. Phys. Ques.}{16}{1914}{719}

\bibitem{hillas} A.M. Hillas, {\em Cosmic Rays,} Pergamon Press, New York 1972
%
%
%
%

\end{thebibliography}
\end{document}